\documentclass{emulateapj}
\usepackage{lscape}
\usepackage{graphicx}
\newcommand{\com}[1]{}

\pdfoutput=1

\begin{document}

\title{Mapping the Recent Star Formation History of the Disk of M51}
\author{Catherine Kaleida\altaffilmark{1}, Paul A.\ Scowen\altaffilmark{1}}
\email{kaleida@asu.edu}
\altaffiltext{1}{School of Earth and Space Exploration, Arizona State
  University, PO Box 871404, Tempe, AZ 85287-1404} 

\shorttitle{Stellar Association Mapping in M51} 
\shortauthors{Kaleida \& Scowen}

\accepted{for publication in Astronomical Journal, May 28, 2010}

\begin{abstract}
Using data acquired as part of a unique Hubble Heritage imaging program of broadband colors of the interacting spiral system M51/NGC 5195, we have conducted a photometric study of the stellar associations across the entire disk of the galaxy in order to assess trends in size, luminosity, and local environment associated with recent star formation activity in the system.  Starting with a sample of over 900 potential associations, we have produced color-magnitude and color-color diagrams for the 120 associations that were deemed to be single-aged.  It has been found that main sequence turnoffs are not evident for the vast majority of the stellar associations in our set, potentially due to the overlap of isochronal tracks at the high mass end of the main sequence, and the limited depth of our images at the distance of M51. In order to obtain ages for more of our sample, we produced model spectral energy distributions (SEDs) to fit to the data from the GALEXEV simple stellar population (SSP) models of Bruzual and Charlot (2003).  These SEDs can be used to determine age, size, mass, metallicity, and dust content of each association via a simple $\chi^{2}$ minimization to each association's $B$, $V$,  and $I$-band fluxes.  The derived association properties are mapped as a function of location, and recent trends in star formation history of the galaxy are explored in light of these results.  This work is the first phase in a program that will compare these stellar systems with their environments using ultraviolet data from GALEX and infrared data from Spitzer, and ultimately we plan to apply the same stellar population mapping methodology to other nearby face-on spiral galaxies.
\end{abstract}


\section{Introduction}
M51(also known as NGC 5194) is one of the nearest and largest grand design spiral galaxies.  The galaxy is known to be undergoing an interaction with its companion NGC 5195, and shows an active system of star formation (SF), largely confined to two spiral arms.  The morphology of NGC 5194 is listed by de Vaucouleurs 1976 as a SA(s)bc(pec)III galaxy.  The companion has been determined to be an SB0 (Sandage \& Tammann 1987), and as an I0(pec) by de Vaucouleurs, and its nucleus hosts a nonthermal source (Ford et al. 1985).  Physical parameters adopted in this paper include a distance of 8.4 $\pm$ 0.6 Mpc (Feldmeier, Ciardullo, \& Jacoby 1997), a tilt angle of 20$\degr$ and a positional angle of the major axis of 170$\degr$ (Tully 1974).  M51 is consistent with super-solar metallicity ($Z \sim$ 2-3 $Z_{\odot}$) and shows a weak metallicity gradient as a function of distance from the nucleus (Zaritsky et al. 1994).  

\subsection{M51 Dynamics}
The M51 system is a very complicated and dynamic system.  Most of the unique properties exhibited by the system appear, not surprisingly, to be directly related to the interaction the disk is experiencing with the companion. This interaction has been modelled extensively throughout the years, with increasing success in matching the present state of the M51/NGC 5195 system.  Early models fit optical data well, but were unable to explain M51's extended HI tail (Toomre \& Toomre 1972; Hernquist 1990; Rots et al. 1990).  More recent attempts fit HI maps well, but differ in the histories they paint for the M51 system.  Salo \& Laurikainen's (2000) N-body model suggests elliptical orbits and multiple passages of NGC 5195 through the disk of M51.  In contrast, Wahde and Donner's (2001) model utilizing a genetic algorithm favors a hyperbolic and thus single-passage encounter, but fails to match the observed velocities in the HI tail.  Theis and Spinneker (2003) combine N-body methods with a genetic algorithm, and achieve a good fit to both the HI intensity and velocity maps.  Their results suggest a highly elliptic orbit with two recent disk passages at 400-500 Myr and 50-100 Myr ago, in agreement with Salo \& Laurikainen. 

\subsection{Star Clusters and Associations in M51}
In addition to its intrinsic properties, the uniqueness of the M51 system as a test bed for understanding the effects of galaxy interaction on star formation has led to many systematic studies of its stellar populations.  Previous HST studies of the compact star clusters in M51 include Bik et al. 2003 and Bastian et al. 2005a,b, who investigated clusters near the center of M51, Lee et al. 2005, who utilized a morphological cluster selection criteria across the whole of the galaxy, and Hwang \& Lee 2008, who detected over 2,000 clusters in the M51/NGC 5195 system.

Observations have shown that M51 exhibits a much bluer color than its companion overall, and the blue and yellow-red stellar populations within M51 are spatially separate structures (Zwicky 1955).  The bluer star cluster colors exhibited in late-type galaxies such as M51 indicate recent star formation, as the O and B stars that produce this light are short-lived.  In addition, interacting galaxies tend to contain very luminous and massive young star clusters, as compared to solitary galaxies (Whitmore 1999, and references therein).  In this paper, and subsequent work, we intend to use the data from the ACS/WFC 10452 imaging survey of M51 to study in detail the structure, dynamics, stellar populations and ongoing star formation in this system.

The high spatial resolution and broad wavelength coverage of the ACS/WFC provide unique data on the stellar populations in and around HII regions in M51.  Our initial objective is to observe colors and magnitudes of stellar associations covering a range in luminosity and galactocentric radius within M51. These data can be used to build color-magnitude (CM) and color-color (CC) diagrams for the youngest stars, and thereby assess the fundamental parameters of the young stellar populations. At the distance of M51, we calculate that the limiting magnitude of the ACS-WFC camera (M$_V$=-3.12) will restrict detection of CM turnoffs to only the youngest and brightest stellar populations, corresponding to an age of $\sim$12 million years in the $V$-band.  Thus, the stellar groupings which we will explore here are generally OB associations ($\sim$100 pc in size) or subgroups within OB associations ($\sim$50 pc) (Brown et al. 1999, Elmegreen et al. 2006).  The stellar data can also be used to assess possible systematic variations in the stellar content of these complexes as a function of position within the galaxy. 

The timescale between local compression and the peak in SF can be made based on the apparent separation of the arms delineated by the dust and HII regions respectively (Shu et al. 1972).  This timescale is found to be a constant at 4.6 Myr for the southern arm (Elmegreen et al. 1992), and increasing as a function of radius for the northern arm, although the actual value for the northern arm timescale is less than that exhibited by the southern arm. By looking at the stellar components across the divide and their apparent ages we can learn how long the youngest stars reside where they were born, and thus track the position of the stars over time and when and where the SF event actually happened. In addition, we can use this stellar association sample to gauge the SF history as a function of position and determine the role the local environment plays in the rate and efficiency of the process.  We already know that environment plays a key role in determining the stability of the local ISM to triggers or compressions that may induce gravitational collapse, and ultimately cause star formation.  What is less clear is how that characteristic is set by the presence of a first generation of stars from an earlier SF event.  Clearly, if a massive stellar cluster is formed, that cluster imparts not only strong UV radiation to the local ISM (as the resulting HII region is not typically radiation bounded), but also leaks strong continuum photons into the surrounding gas.  The cluster can also contribute strong mechanical energy from its intense stellar winds that can sweep up and collect material into dense sheets or interfaces.  These interfaces are ripe for the second generation of SF to ignite upon the incidence of a supernova blast wave (from the same short-lived massive stars), or from the nearby compression of material due to a variety of other possible triggers, such as cloud-cloud collisions, spiral density waves, etc..

\vspace{10mm}
\section{Observations}

The datasets used in this study were obtained with the Hubble Advanced Camera for Surveys/Wide Field Camera (ACS/WFC) as part of Hubble Space Telescope (HST) program 10452, which was completed in January 2005 and became publicly available in April 2005 (Mutchler et al. 2005).  The datasets provide a six-tile image mosaic, with four dithered exposures taken for each of four filters.  The 4 different imaging bands are $B$, $V$, and $I$ broadbands (F435W, F555W and F814W respectively), and the narrowband redshifted $H\alpha$ filter F658N. The standard archival data was retrieved ``on-the-fly'' from the archive with standard ACS pipeline (CALACS) processing, including bias, dark, and flat-field corrections. The $H\alpha$ exposures are of sufficient depth to reach surface brightnesses of at least 10$^{-16}$ ergs/cm$^{2}$/s/arcsec$^{2}$ to allow us to probe the faintest HII regions.  

An ACS/WFC pixel subtends 2.0 pc at the distance of M51 and its companion NGC 5195. This means that a single pixel in the image could contain one or multiple stars, and thus there is the possibility of more than one star being sampled by each pixel. If two or more stars fall on the same pixel, this will still yield a non-symmetric wing in the PSF that may be detected in the surrounding pixels.  Also, the total luminosity of a pixel is a good determinant of the number of stars in a single pixel or source. If the intrinsic brightness of a pixel or source is greater than the maximum luminosity expected for a single star then it can be assumed that the pixel contains multiple stars.  On our CM diagrams, luminosities at or above $M_{V}$=$\sim$-9 are assumed to be multiple stars or compact clusters. 

\vspace{-4mm}
\subsection{Pointings} 

Using the Wide Field Channel (WFC) of the Advanced Camera for Surveys (ACS), a 2$\times$3 mosaic (6 pointings, 4 orbits per pointing) was obtained in 4 filters (Johnson $B$,$V$,$I$, and $H\alpha$) at a telescope orient of $\sim$270$\degr$.  Details of these exposures along with the central coordinates of the 6 pointings or ``tiles" in the 2$\times$3 mosaic are available in Mutchler et al. 2005.  Each pointing provides 4 orbits of integration providing the limiting magnitudes listed in Table 1.  The 4 exposures within each tile were dithered: a small sub-pixel dither (2.5$\times$1.5 pixels), and a larger dither which spans the interchip gap (5$\times$60 pixels).

\section{Data Reduction}

\subsection{Processing}

The initial raw data was processed using the standard Hubble pipeline, through the IRAF task CALACS (Mutchler et al. 2005).  CALACS performs the corrections for bias, dark, and flat fielding. The resulting images were then cosmic-ray rejected by combining two images of the same exposure time and gain, discarding the saturated cosmic ray pixels from one image, and replacing the ``bad" pixel with the good signal from the second image. Distortion effects from the camera's off-axis position within the observatory are smaller than the photometry aperture used, and thus are inconsequential for this application (but would be relevant if exact astrometry of individual stars was needed). 

\subsection{Photometry and Initial Stellar Association Selection}

    The associations included in this catalog were chosen to be relatively isolated from neighboring groups of stars in order to produce a single-aged CM diagram. After processing with CALACS and cosmic-ray rejection, the V-band image was inspected by eye for isolated stellar groups from which CM and CC diagrams could be derived.  Initial sample selection was performed by visually scanning each image for groupings of stars/sources that could be associated with one another.  A total of 969 stellar associations were selected in this way for subsequent analysis.  For each association, a CM diagram was produced to assist in determining if the group of stars in question were indeed coeval.  A more detailed description of this method of analysis is described in following sections.  Photometry was performed on these groupings of stars using the GSFC IDL ASTROLIB package IDLPHOT.  This package selects the individual stars and outputs the flux from each star, as measured within an aperture with a two pixel radius, in the $B$, $V$, and $I$ broadband filters and the $H\alpha$ narrowband filter. Two pixels is the approximate full width at half maximum (FWHM) of point sources in these images, and this aperture size was chosen to help exclude unresolved compact clusters from our source selection, although selection of the most compact clusters along with the stars is unavoidable. Thus the ``sources'' plotted include stars, unresolved multiple star systems, and some compact clusters.  Background maps for each tile, chip and imaging band were produced by iteratively replacing any pixel greater than 1.3 times the median of the surrounding 12x12 pixel box with the median value, until all stellar sources were masked out by the median value.  These background maps were then subtracted from the images to properly correct for the local sky level.  The fluxes derived from our initial methods were then converted to absolute Vega magnitudes by dividing by the exposure time, multiplying by the spectral flux density per unit wavelength that would generate 1 count/sec (header value PHOTFLAM), and then translated to an $ST$ magnitude via the equation m = -2.5 log$_{10}$(F) + PHOTZPT (where PHOTZPT=-21.10 for the $ST$ magnitude scale). These magnitudes were converted to Vega magnitudes and compared to magnitudes calculated via the method detailed in Sirianni et al. (2005). The magnitudes calculated in these two methods were identical to within a thousandth of a magnitude.  Finally, a CM diagram of absolute magnitude in the $V$-band versus ($B-V$) and a CC diagram of ($B-V$) versus ($V-I$) were produced for each stellar association.  

\subsection{Extinction/Reddening and Redshift Corrections}

Photons collected from the M51/NGC 5195 system experience reddening and extinction due to dust internal to M51 and Milky Way dust, as well as redshift from the radial component of the galaxy's proper motion.  All of these factors must be carefully corrected to insure accurate photometry.  The redshift of M51 is z=0.00154, which corresponds to shifts of 6.62$\AA$, 8.23$\AA$, 12.8$\AA$, and 10.1$\AA$, for the central wavelengths of the $B$, $V$, $I$, and $H\alpha$ filters, respectively (Scoville et al. 2001).  These corrections were applied to all wavelengths.  Extinction due to dust poses a more complex problem, as decomposing internal and Galactic dust is extremely difficult without data across the whole spectral range.  Also, most authors utilize the Galactic extinction corrections from Table 6 of Schlegel et al. 1998, but these values are not ideal for our purposes since they were acquired by convolving the SED of an elliptical galaxy with the specified filter throughput curves.  Assuming the SED of an elliptical galaxy neglects the blue continuum from young stars present in a spiral galaxy, and would introduce an $\sim$5$\%$ error in our reddening corrections (Schlegel 2008, personal correspondence). For the present analysis, we treat the internal and Galactic extinction separately, and follow the precedent of other authors (Lee et al. 2005, Mora et al. 2009, Ubeda et al. 2007) and use the Schlegel et al. 1998 values to correct for extinction due to dust in the Milky Way, in spite of the 5$\%$ error.  Future work is planned to explore the possibility of avoiding this error by producing de-reddened SSP SEDs from the GALAXEV models and fitting the reddening values to the data available from Hubble, GALEX, and Spitzer for each stellar association.  This method would bypass the necessity of decomposing internal and Galactic reddening, and avoid the error imposed by assuming the SED of an elliptical galaxy.  For the present study, we correct for Galactic extinction using equations 1, 3a, and 3b in Cardelli, Clayton, and Mathis 1989, with $R_{V}$=3.1 and A$_{V_{MW}}$=0.117 for the direction of M51 (Cardelli et al. 1989, Schlegel et al. 1998).  

The intrinsically dusty environment of M51 is known to hamper studies of both the ionized gas and the HII regions themselves.  The overall apparent structure and excitation state of gas around massive ionizing stars is strongly affected by the presence of the dust and may modify the continuum of photons that escape to ionize and energize the surrounding interstellar medium (ISM).  The effect these modifications have on the dynamics and susceptibility of the ISM to subsequent second-generation triggering or propagation of star formation make this galaxy an interesting candidate for the study of star formation as a global mechanism for distributing and regulating the formation of the next generations of stars. To account for the effects of dust internal to M51, we assume a global extinction correction of A$_{V_{M51}}$=0.3 mags, corresponding to an reddening correction of E(B-V)=0.097, using R$_V$=3.1 (Cardelli et al. 1989). This value was found to be an average of the best fit for the majority of the associations studied, if A$_{V_{M51}}$ is allowed to vary between 0-1 mags, in increments of 0.1 mags.  This value is in good agreement with with Hwang \& Lee 2008, who use E(B-V)=0.1 (A$_{V_{M51}}$=0.31 mags), and is less than the maximum value in the bulge, which is E(B-V)=0.25 (A$_{V_{M51}}$=0.775) (Lamers et al. 2002).  Ideally, it would be best to leave the extinction correction as a free parameter in the SED fitting, since dust content varies noticeably over the disk of M51, and thus a single correction value for the entire galaxy can only be an approximate correction.  Unfortunately, with only $BVI$ fluxes, we are unable to discern fluctuations in the local dust environment and stellar association ages simultaneously.  Visual inspection of the CM and CC diagrams of the 120 associations in our set show that A$_{V_{M51}}$ lies between 0 and 1 magnitudes for most of the associations.  We choose an intermediate value of 0.3 magnitudes, knowing that some associations will be slightly over- or under-corrected.  Once redshift corrections and internal and Galactic reddening values.
 
\subsection{Stellar Association Selection and Age Determination}

The use of a CM diagram of M$_{V}$ versus (B-V) was originally chosen to allow direct comparison with the Sandage age estimation model (Sandage 1957), in order to test the viability of determining stellar association ages via their main sequence (MS) turnoff magnitude, given the distance of M51 and depth of HST ACS/WFC images.  These colors were chosen over other available bands to maximize the depth of the sample, as $B$ and $V$ have fainter limiting magnitudes than $I$ or $H\alpha$.  The Sandage model is based on MS turnoff ages of 10 galactic clusters and one globular cluster.  These clusters provide an age-MS turnoff magnitude relation, which we derived via a simple empirical interpolation of Sandage's Figure 1.  For each association in our sample, a selection box of the appropriate size was chosen to produce the narrowest sequence and thus the tightest fit to the Sandage model when it was overlaid on the resulting CM diagram.  Dense stellar regions which did not closely fit the model were not included in the list of potential associations, as they are most likely blends of multiple cluster.  These groups may be spatially related, but are not necessarily coeval, making determination of a single age impossible.

More modern treatments of this approach involve the use of model isochrones to accurately predict the apparent locus of stellar points on such a CM diagram for known differences in metallicity, age, and cluster size.  However the underlying treatment is still sound - to use the distribution of member stars in an association to determine an estimate for the apparent age of the grouping based on the turnoff exhibited.  Complications arise in the application of the method due to blends of multiple stars and/or compact clusters being detected as a single source.  Plotting sources in both color-magnitude and color-color space helps to separate stars from compact clusters, and to weed out groups of stars that are not massive enough to be accurately fit with the Simple Stellar Populations models of BC03.  The M51/NGC 5195 system is at the very limit of the HST/ACS camera's ability to resolve individual stars, and the corresponding brightness limit due to the exposure times used narrows the ages we could detect via this method to only the most recent of SF events.  Consequently, this study is only sensitive to the bluest of associations containing the most massive, and therefore the shortest-lived, stars. 

Even if blending or resolution limits were not an issue, the fact that we are working at the blue end of the MS also constrains the resolution in magnitude space that can be used to resolve turnoffs on CM diagrams.  At the top end of the MS, the sequence is almost vertical, meaning that the locus of turnoff points is also almost vertical.  The degree to which we can measure small changes in brightness therefore limits the degree to which we can pull out departures from the MS due to evolutionary turnoff, as opposed to a simple intrinsic spread in brightness due to population and possibly internal reddening etc.  There is an intrinsic scatter in apparent brightness for any collection of stars in a cluster or association, and it is the art of distinguishing intrinsic scatter from changes due to stellar evolution that is at the heart of this approach, and is also the most difficult part of the technique.  The limiting magnitude of the ACS observations translates into an inability to detect main sequence turnoff ages any older than $\sim$12 million years.  This is a flaw in the approach, but given the distance to the galaxy, this is at the boundary of what we can do observationally at this time (Rey et al. 2007). 

To help make the final selection of single-aged associations to be included in the sample, Padova stellar isochrones (Girardi 2006, http://pleiadi.pd.astro.it/isoc$\textunderscore$photsys.02/isoc$\textunderscore$acs$\textunderscore$wfc/) and a BC03 stellar population evolutionary track were overlaid on the CM and CC diagrams.  Figure 1 shows these plots for three example associations from our set that span the range of ages found (4 to 610 Myr).  One must carefully assess the distributions of sources for each association on these two plots to be certian that the stars/compact clusters in it are in fact coeval.  Groups of stars that did not lie along a single isochrone (within the scatter of the data), whose total magnitude did not align with the BC03 model SED, or whose CM and CC diagrams indicated discrepant ages were excluded from the set.   This second iteration of the selection process yielded a final list of 120 associations.  Figure 1 shows the CM and CC diagrams for three representative stellar associations from this set, alongside the image subraster that generated it. 

\noindent
\begin{figure*}[htbp]
\begin{center}
\includegraphics[scale=0.8]{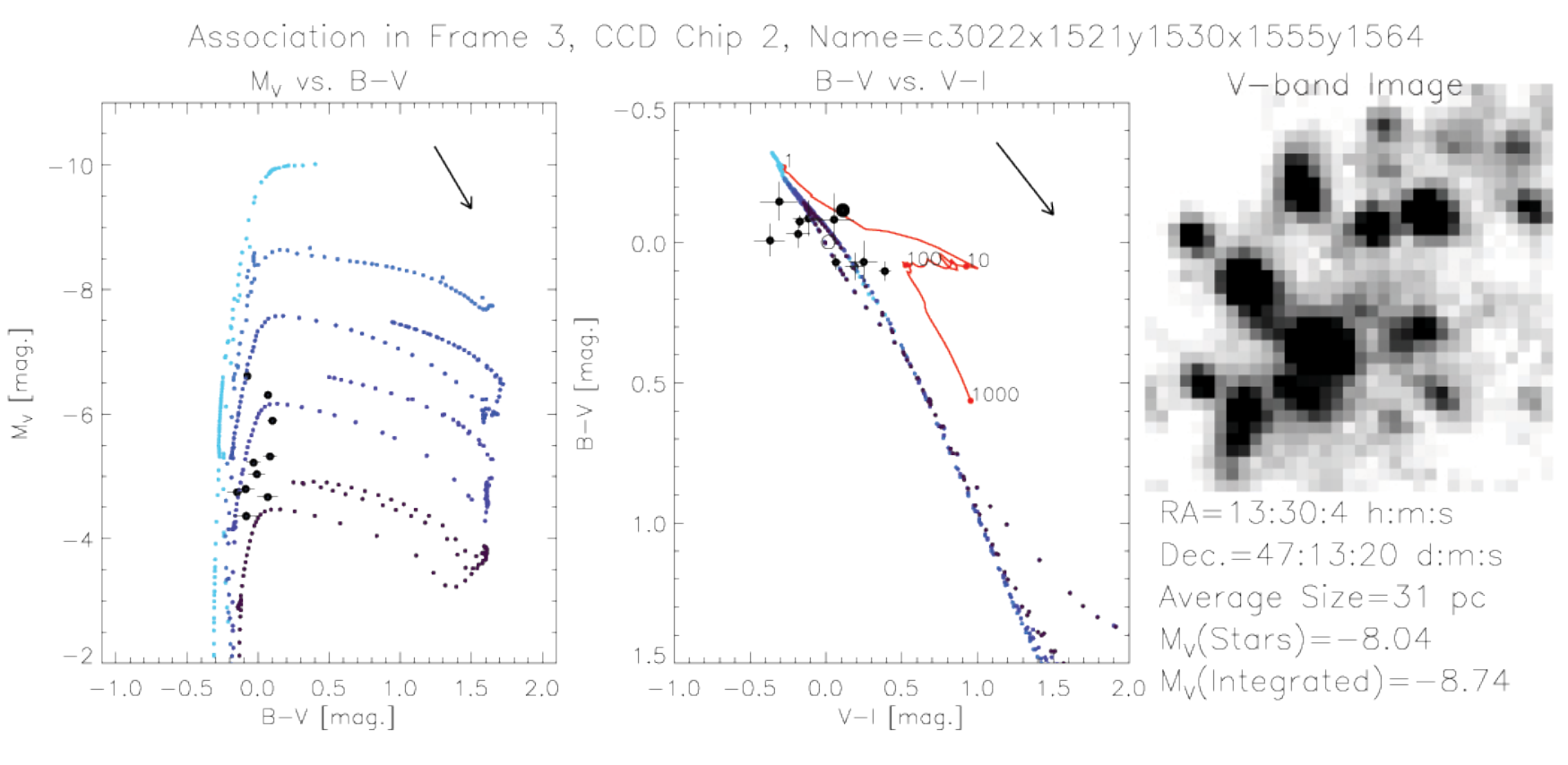}\\
\vspace{3mm}
\includegraphics[scale=0.8]{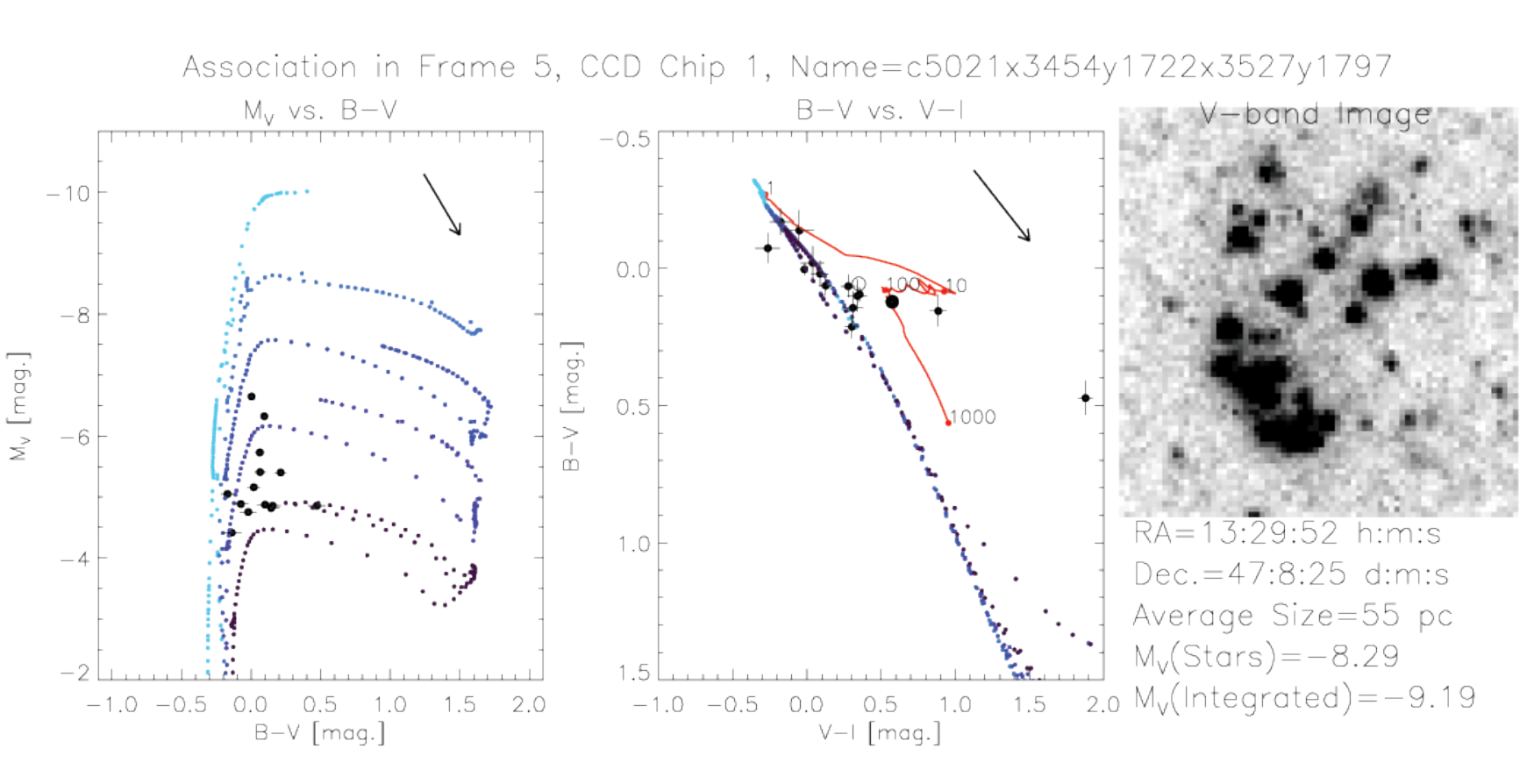}\\
\vspace{3mm}
\includegraphics[scale=0.8]{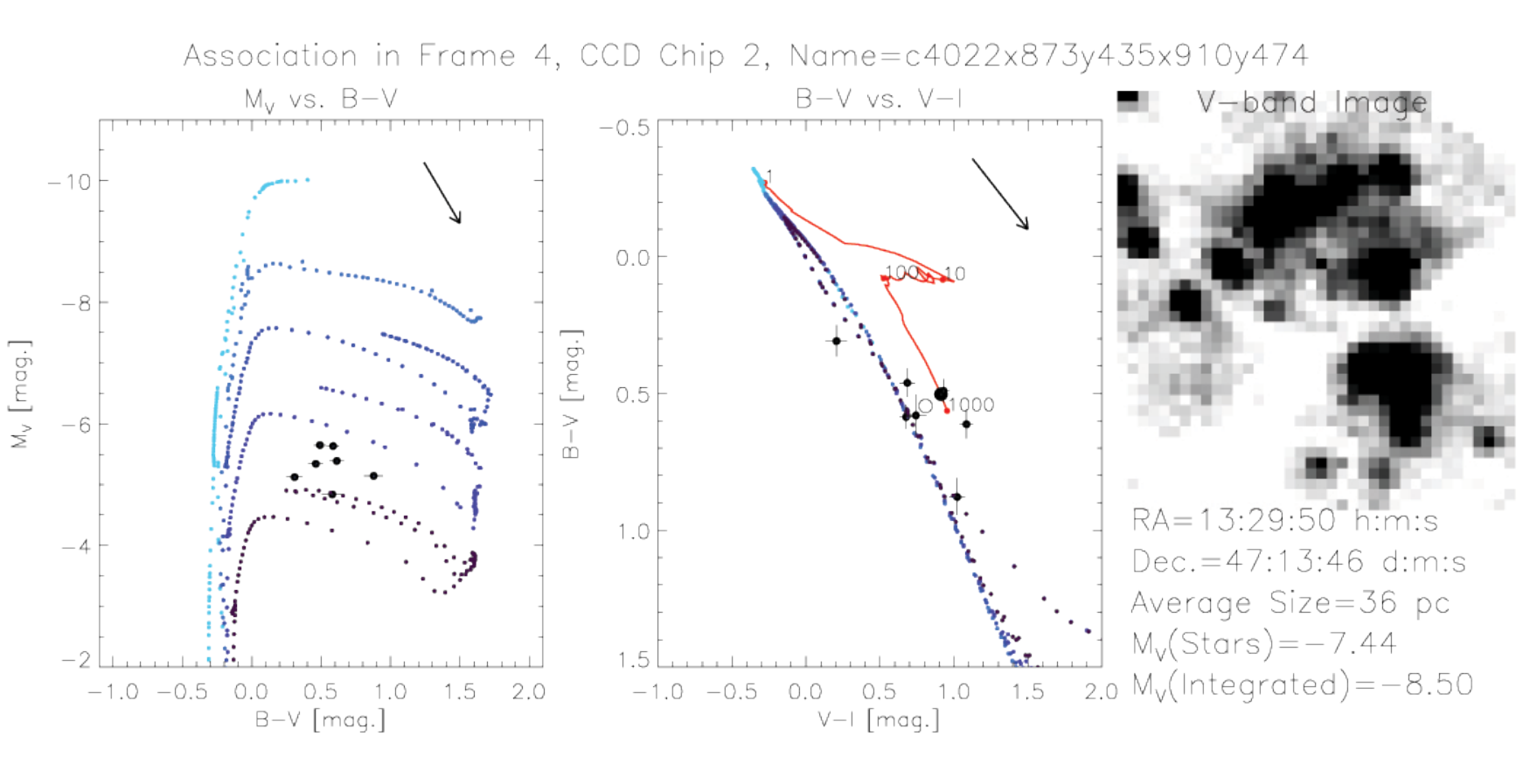}\\
\footnotesize
\caption{CM and CC diagrams, and $V$-band images for three example stellar associations from our set.  The blue-purple dotted lines are Padova stellar isochrones for stellar masses from 0.15 to 66.56 MSun, metallicity Z=0.019, and ages are 1 Myr (lightest blue), 6 Myr, 10 Myr, 20 Myr, and 50 Myr (purple) (Girardi 2006, 
http://pleiadi.pd.astro.it/isoc$\textunderscore$photsys.02/isoc$\textunderscore$acs$\textunderscore$wfc/).   The red line in the left hand panel is a BC03 model evolutionary track for an instantaneous burst stellar population of metallicity Z=0.02.  Datapoints have been corrected for Milky Way dust, and A$_{V_{M51}}$=0.3 mags has been added to all models plotted.  The two large dots represent the total association flux from summing the sources (open circle), and the total flux from within a large circular aperture centered on the association(closed circle).  Ages for these three associations are, from top to bottom, 4 Myr, 160 Myr, and 610 Myr.}
\end{center}
\end{figure*}

\vspace{-\baselineskip}
Since M51 is at the limits of our ability to resolve individual stars and measure MS turnoff ages, we decided to employ a modern refinement - the use of stellar population models in tandem with SED fitting to allow the prediction and subsequent measurement of stellar association age.  In order to fit model SEDs to the selected stellar associations, we have written a series of IDL scripts that calculate the flux from each of the associations, produce model stellar population SEDs from the GALAXEV library, calibrate these models for comparison to the data, and perform a $\chi^{2}$ fit between the observations and the model SEDs.  The SED with the minimum $\chi^{2}$ value when compared to the data is the best-fit model.

The GALAXEV library allows for tuning of the input parameters in order to best match the stellar population of interest, in our case stellar associations in the M51/NGC 5195 system.  For associations in M51 we choose the Padova 1994 stellar evolutionary tracks (BC03, Alongi et al. 1993, Bressan et al. 1993, Fagotto et al. 1994a,b, Girardi et al. 1996), a Salpeter initial mass function of $\xi(log_{10}(M)) \propto M^{-1.35}$ (Salpeter 1955), and an instantaneous burst SFR of $\psi(t)=1 M_{\odot} \cdot \delta(t)$ (BC03), which best replicates the scenario of triggered star formation.  We allow for a range of models between the ages of 0.1 Myr to 1 Gyr (this range should include the two most recent perigalacticon passages of the two galaxies) (Salo \& Laurikainen 2000), extinction due to dust in M51 of $A_{V}$=0.3 magnitudes, and a metallicity of 2.5$Z_{\odot}$ (Zaritsky et al. 1994). The mass of each association is quantitatively calculated during the $\chi^{2}$ fitting process, details of which are described below.

To perform model SED fitting, a library of BC03 Simple Stellar Population models was produced, and the model stellar association SEDs were created by converting the GALAXEV output into fluxes that can be compared directly to fluxes calculated from the Hubble data. The library of BC03 SSP models are created by running two scripts available in the GALAXEV package: CSP\_GALAXEV and GALAXEVPL.  CSP\_GALAXEV evolves the composite stellar populations spectrally for a constant metallicity using equation (1) of Bruzual \& Charlot 2003, and outputs a binary file of the evolved spectral evolutionary model.  The program GALAXEVPL then reads in the binary file from CSP\_GALAXEV, and outputs ASCII files of SEDs at the ages requested by the user. This ASCII file contains wavelength (column 1) and specific luminosity density ($\ell_{\lambda}$ in the Papovich 2001), or luminosity per unit wavelength, in units of $L_{\odot} \cdot \AA^{-1}$ (column 2).  In order to convert from this number to flux ($F_{\nu}$), we followed the procedure outlined in Section 4.2 of Papovich et al. 2001, using 
\begin{equation}
L_{\nu} = \frac{10^{8} \lambda_{0}^{2} \ell_{\lambda} L_{\odot}}{cM_{tot}} 10^{-0.4(A_{\lambda,MW}+A_{\lambda,M51})}
\end{equation}
to calculate the luminosity per unit frequency, and 
\begin{equation}
F_{\nu} = (1+z)L_{\nu}\frac{1}{4\pi d^{2}}
\end{equation}
to calculate the flux per unit frequency, where $\lambda_{0}$ is the redshifted wavelength, $\ell_{\lambda}$ is the specific luminosity density, $M_{tot}$ is the total mass of the stellar population at a particular timestep, $A_{\lambda,MW}$ is the extinction due to dust in the Milky Way (Cardelli et al. 1989, Schlegel et al. 1998), $A_{\lambda,M51}$ is the extinction due to dust internal to M51, d is the distance to the galaxy, and all units are in the CGS system.  $F_{\nu}$ calculated from equations (1) and (2) is for a 1 $M_{\odot}$ stellar population, and has units of erg $\cdot$ cm$^{-2}$ $\cdot$ s$^{-1}$ $\cdot$ Hz$^{-1}$.  The quantity $M_{tot}$ is obtained by extracting the value of Mgalaxy in column 9 of the *.4color ASCII file output from GALAXEVPL which corresponds to the desired age to be modelled, and is the same quantity as $m^{*}(t)$ in Papovich et al. 2001.

We then convolved each model SED with the $B$, $V$, $I$, and $H\alpha$ filter throughput curves to get the flux per unit frequency in each filter, $f_{\nu}$, using 
\begin{equation}
f_{\nu}=\frac{\int \frac{F_{\nu} \tau_{\nu} }{\nu}d\nu}{\int \frac{\tau_{\nu} }{\nu}d\nu}
\end{equation}
where $\tau_{\nu}$ is the total system throughput curve, including all transmission and detection effects of the telescope, camera, filter, and CCD chip (Fukugita et al. 1996).

The mass of each stellar association is calculated using the fact that the BC03 model SEDs are normalized to a 1 $M_{\odot}$ stellar population.  The mass which produces the smallest $\chi^{2}$ value between the data and each model SED is calculated by minimizing the following equation with respect to $M$ and then solving for $M$, then plugging this mass back into the equation to calculate the minimum $\chi^{2}$ for each model.
\begin{equation}
\chi^{2}=\sum_{i} \left[\frac{f_{\nu,i}^{obs}-Mf_{\nu,i}}{\sigma_{i}}\right]^{2}
\end{equation}
where the summation is over $i$ filters, $f_{\nu,i}^{obs}$ is the observed flux in filter $i$, $M$ is the best-fit mass for a particular stellar association, $f_{\nu,i}$ is the model SED flux in the filter $i$, and $\sigma_{i}$ is the error in the observed flux for filter $i$.  The best-fit model stellar association SED was then chosen from the array of model SEDS by finding the minimum $\chi^{2}$ value between model and data in the $B$, $V$, and $I$ bands.  It should be noted that it is not valid to use $H\alpha$ flux directly as a fitting parameter for a stellar population, since $H\alpha$ is emitted and absorbed both from the stellar photosphere and the surrounding gas.

\subsection{Other Parameters (Stellar Association Size, Position)}

The size of each association was estimated by determining the spatial extent of the full width at half maximum (FWHM) of the flux density profile of stars.  Specifically, stars/sources were counted in radial bins from the association center, the flux from sources in each bin were totaled, and the standard deviation of the spatial flux density distribution was computed.  The full width at half maximum was then computed assuming that the distribution was Gaussian.

The position of each stellar association in equatorial sky coordinates was calculated using the WCSTools package (Mink 1997, 1999), which allows for the non-linear conversion from pixel coordinates in the undrizzled images output from CALACS to right ascension and declination in the appropriate epoch.  Associations were named according to the tile of the mosaic they reside in, and the lower left and upper right corners of the bounding box for that association, in pixels. 

\section{Results}

\subsection{Properties of Isolated Stellar Associations in M51}

Results presented here are from SED fitting of the ages, masses, metallicities, and internal extinctions for each stellar association. Ages were also assessed by attempting to pin down the location of the main sequence turnoff point on the color-magnitude diagram, but we found that turnoffs were not apparent for the vast majority of associations (see Discussion).  Table 1 lists the derived properties for the 120 associations selected from our sample using the criteria and methods laid out above.  The identifier for each association is based on a 4-digit number indicating which image in the mosaic the association is located in, and the x,y pixel coordinates of the lower left and upper right corners of the selection box used for selection of that association. The location of the central pixel in Epoch 2000.0 coordinates and BVI magnitudes in the Vega system are listed for each association.

It should be emphasized that this is the subset of associations from the full sample whose CM diagrams indicated that they were likely to be single-aged, allowing indisputable measurement of the properties listed, within the limitations of the method.  

Figure 2 shows a map of the galaxy with various annotations derived from the measurement exercise described above, and a number density histogram of the stellar association size sizes.  On this diagram each of the associations has been pictorially represented by a colored circle.  The size of the circle reflects the relative size of the association in physical extent.  The color of the circle depicts the best-fit age of the association in question - with red associations being the oldest at $\sim$600 million years, ranging down through the rainbow to the youngest ages found at 4 million years, in blue.  Figure 3 shows a similar map of the galaxy, where the size of the circle corresponds to the best-fit mass calculated quantitatively from the SED model, instead of the association size.  Masses found range from $\sim$1.1$\times$10$^{3}$-7.4$\times$10$^{6}$ $M_{\odot}$.  The contents of Table 1 are therefore pictorially represented in these two galaxy maps.

%Figure 2
\begin{figure*}[htbp]
\begin{center}
\includegraphics[scale=1.4]{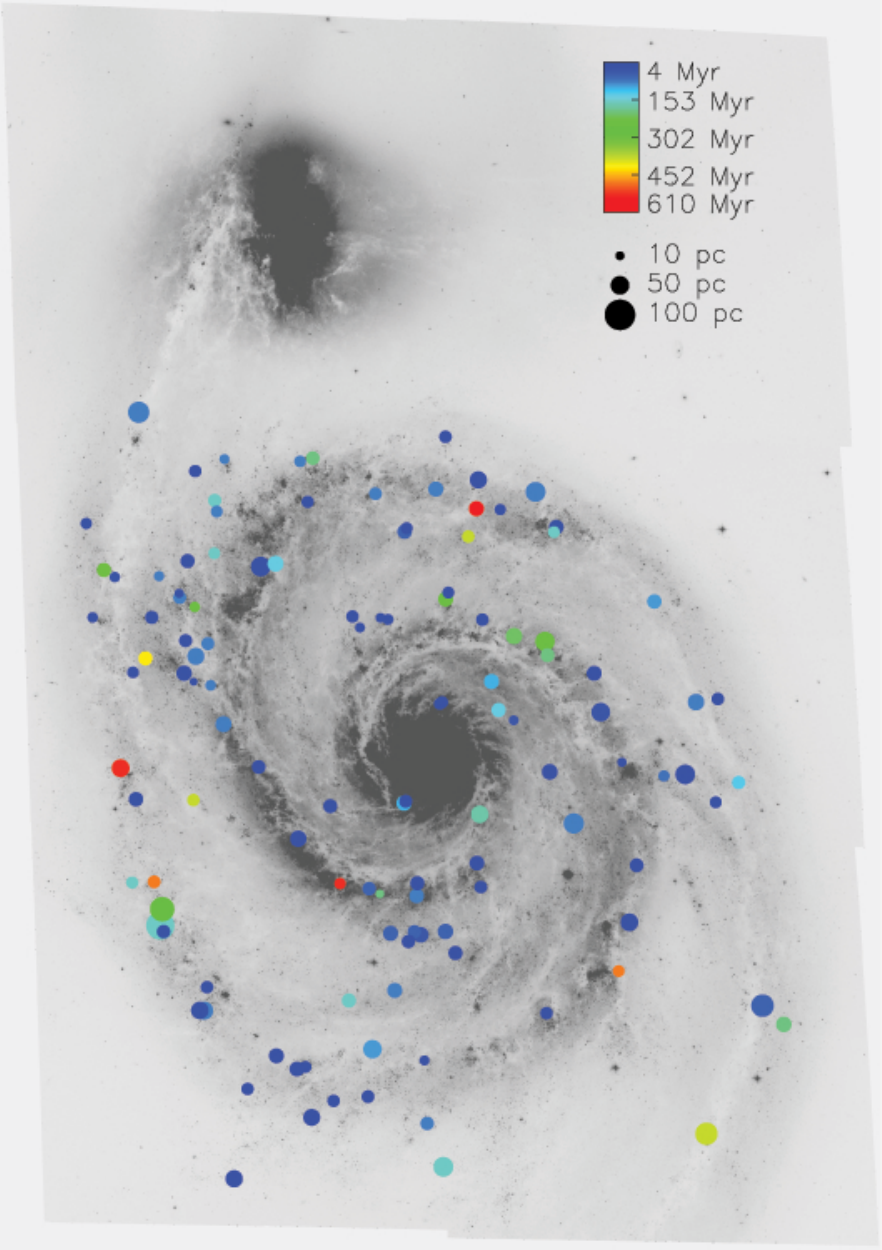} \\
\includegraphics[scale=0.5]{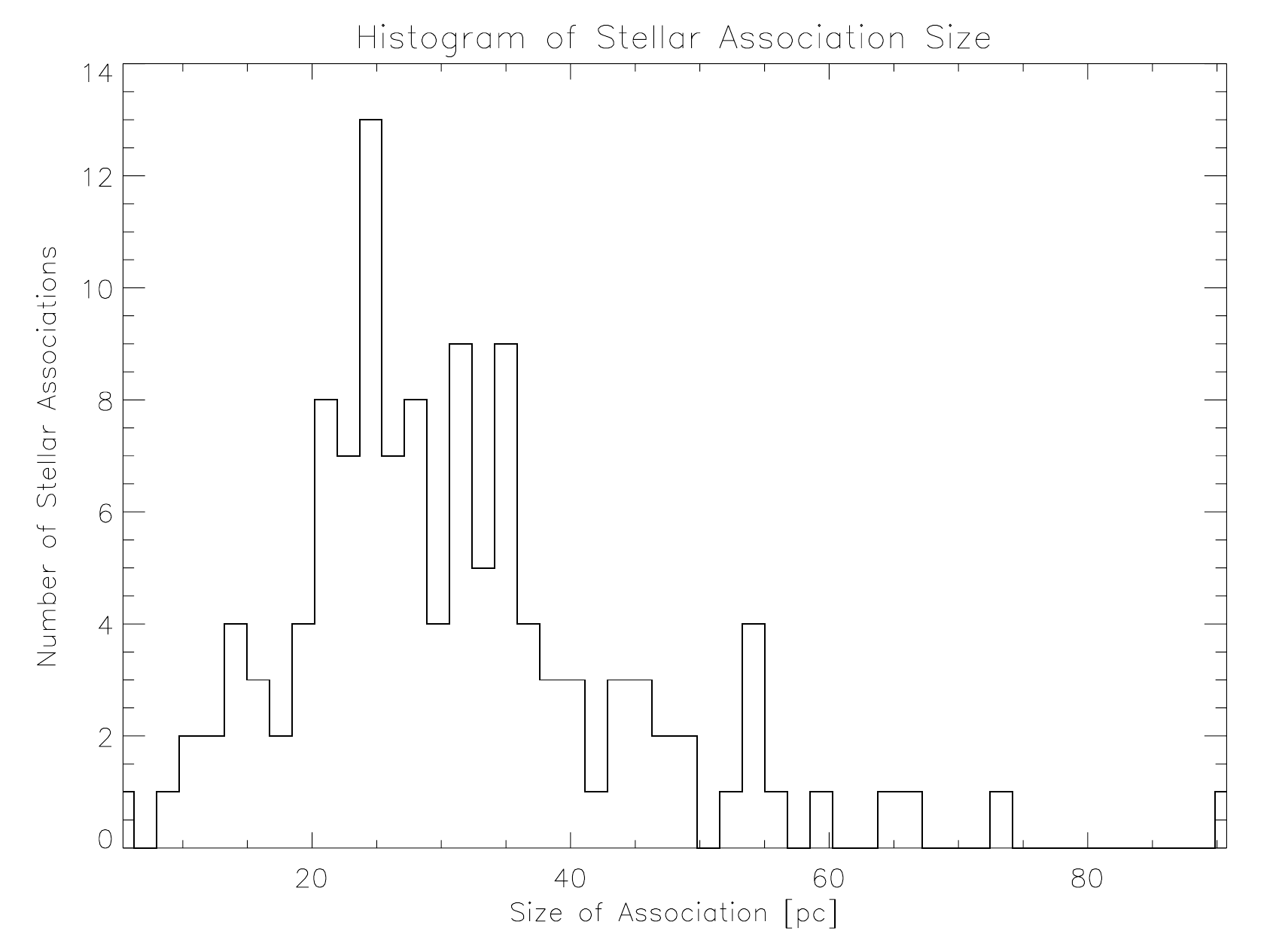}
\footnotesize
\caption{Top: A map of M51 indicating the age, size and location of each of the associations in our 120-association sample.  For each association the relative size is represented by the size of the circle used, the age is represented by the color of the circle, and the location is presented as an overlay on the full field ACS V-band image itself. Bottom: A histogram of number of associations per size bin.}
\end{center}
\end{figure*}

%Figure 3
\begin{figure*}[htbp]
\begin{center}
\includegraphics[scale=1.8]{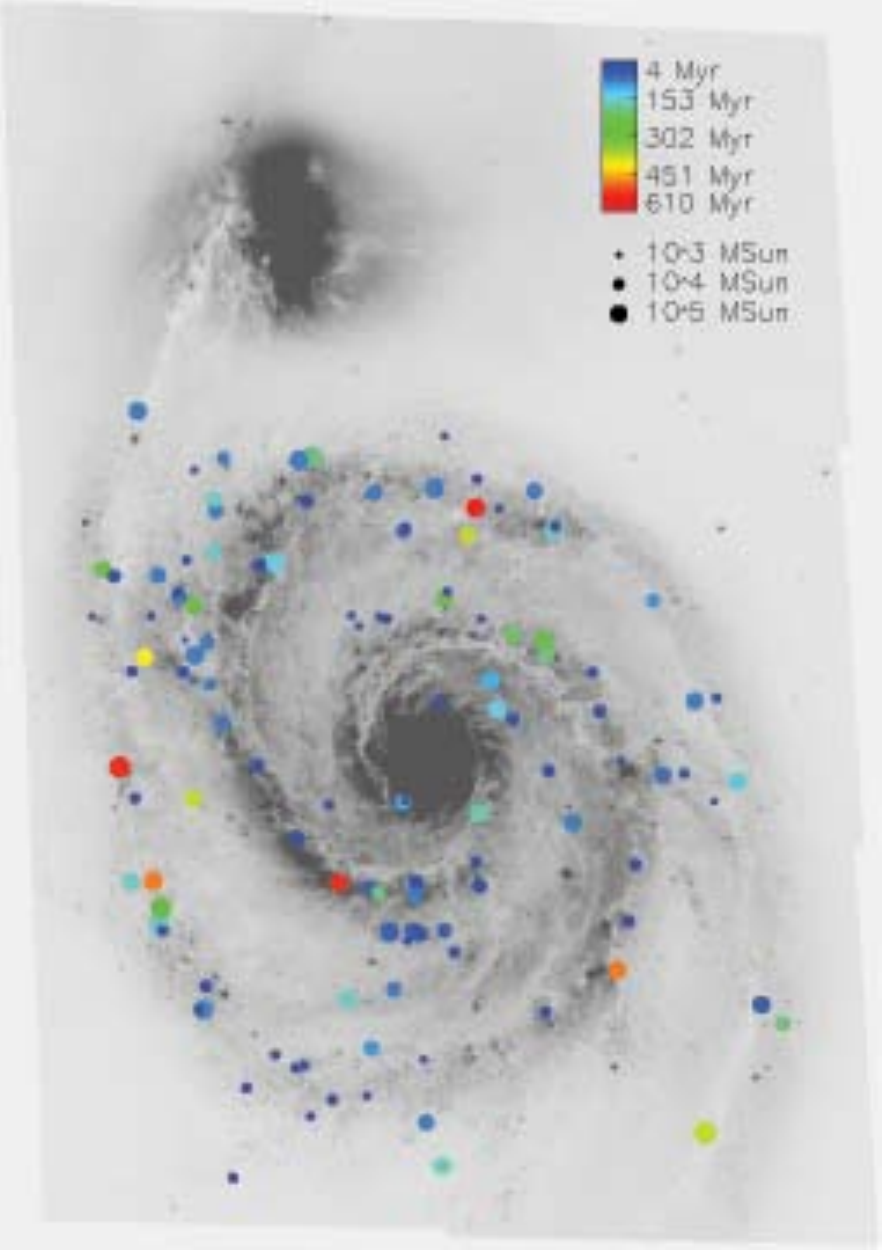}
\footnotesize
\caption{A map of M51 indicating the age, mass and location of the 120 associations in our sample.  For each association the mass is represented by the size of the circle used, the age is represented by the color of the circle, and the location is presented as an overlay on the ACS V-band image.}
\end{center}
\end{figure*}

\section{Discussion}

\subsection{Age, Size, Mass, and Dust Distribution with Position}

Examination of Figure 2 (top) shows that the majority of our association sample is resident on one or the other of the two primary spiral arms in the galaxy, or on spurs coming off the spiral arms, with few found in the interarm region. This is not a surprising result, but considering that associations in our set are defined to be single-aged groupings, this makes a statement about the dominant outcome of SF in these regions due to the recent SF episodes active in the disk.  The lack of single-aged groupings in the inter-arm region may indicate that stellar associations disperse relatively quickly after formation.  There is also a noticeable concentration of associations along the northern arm, the arm closest to NGC 5195, indicating an enhancement in the most recent SF activity in that area.  The distribution of stellar association ages along this arm, coupled with the relative spatial positions of M51 and NGC 5195 indicates that gas is actively streaming along the arm and forming stars in its wake.  Clearly, our association selection criteria limits the sample to the most dominant SF regions, since we require the associations to be well defined and well isolated from neighboring associations or dense collections of field stars.  Figure 2 shows no noticeable trend between global position and association age, but does show some local trends indicating the possibility of secondary and tertiary SF events. There is also no apparent correlation between association age and distance from the galactic nucleus measured along each spiral arm. The sequencing of stellar associations spatially with age, which one would expect from SF propagating from a trigger, are apparent in some instances, particularly on the bend of the Northern spiral arm closest to the companion, but is not widespread across the whole of the galaxy.  The cycling time nesessary to go from one generation of stars to the next is $\sim$5 Myr, meaning that many cycles of triggered star formation could have occured in this time range (4 to 610 Myr).  It could be the case that while we see the sequential propagation of triggered star formation on star cluster scales (1-5 pc), this propagation may not occur on the larger stellar association and star complex scales ($\sim$50-200 pc) investigated in this study.  Another interesting result is that we see no apparent surge in the local SFR at either of the two ``breaks'' in the spiral arms, where the gravitational effects of the interaction has warped the spiral structure.  One might expect an increased SFR in these regions, but the lack of associations found here indicates that conditions are not conducive to the condensation of gas into stars.  It is likely that the dynamical shear at these two ``breaks'' disrupts the collapse of H$_{2}$ and HI clouds into stars.  The size histogram shows a peak at $\sim$25 pc, which may be due to the mixing or collapse length in the instellar medium dictating the dominant size of the fragment of the molecular gas cloud that will collapse into a stellar association.

In Figure 3, we see that there is a tendency towards more massive stellar associations in the northeastern side of the galaxy, particularly near the nucleus and on the northern spiral arm.  This trend hints at the fact that stellar association mass may be more sensitive to triggers such as galaxy interactions and spiral density waves than age or size of the association.

The main sequence turnoff age determination method proved to have several major limitations.  The vertical and horizontal location of any apparent MS turnoff is limited by the vertical slope of the MS at this location in the CM diagram, and the horizontal scatter of member stars in the association.  Since our limiting magnitude of M$_{V}$=-3.12 defines that we are only seeing the very upper end of the MS, the OB stellar population. Typical lifetimes of the members of this population are really quite short, and indeed are much shorter than the 12 million year age cutoff that the $M_{v}$=-3.12 should represent.  As such many members of the associations we are studying never achieved turnoff, but instead probably went supernova.  Older associations would, as a direct extension of this principle, be composed of much fainter stars and would therefore fall below both our limiting magnitude and selection criteria for analysis as part of our sample, and not be resolved into individual populations.  Analysis of turnoff tracks based on stellar evolution theory also indicates that those B stars that do survive this long follow much smaller turnoff tracks than we expect to see from lower mass stars further down the MS.  All of these factors combine to diminish the size of the effect we are trying to measure and therefore act to reduce the final sample of associations from which we could draw conclusions that could withstand scrutiny. Technically, the main sequence turnoff method should allow detection of association MS turnoffs up to 12 million years of age, but no associations with a turnoff at this brightness were found,  because of the confusion limit encountered when trying to verify the integrity of a turnoff right at the vertical magnitude limit of the observations on diagrams like the one depicted in Figure 1.

\subsection{The Recent Star Formation History of M51}

Using all the results presented in this paper, we can make several rather broad statements about the recent SF history of M51.  It is clear from Figure 2 that the interaction with NGC 5195 has generated many large, well-isolated young stellar associations across the northern arm of M51, many more than are apparent along the southern arm.  This implies an enhancement in the SF on that side of the galaxy, but does not represent a system-wide elevation in the SF rate (SFR).  Within the limitations of the measurements we have been able to make as part of this study, we cannot see any systematic age sequencing of SF as a function of location.  

This is somewhat disappointing since one would expect to see evidence of age sequencing with location as a direct result of SF propagation in the disk.  Why do we not see such evidence?  Aside from the limits of the data, it is very possible that the ISM in the galaxy is sufficiently well-mixed and in an increased dynamic state that it is possible to trigger SF in any given location far easier than it might be with a more quiescent ISM.  We are encouraged that our sample reflects our original expectations concerning enhanced SF on the northern arm, it would however appear that that SF locations are much more random in nature.

Indeed, the quiescent SF state reflected by the southern arm of the galaxy seems to represent a very low level of active SF over the last 5 million years, when compared to the northern arm.  As noted in the Introduction, the expected timescale between the passage of the compression wave in the spiral arm and the formation of stars is just less than 5 million years, and so this would suggest that these results are commensurate with this view, but that the size of the compression apparent in the northern arm, and therefore the SFR this represents, is dramatically elevated, and would result in a different Schmidt law.  These results explicitly depict a large variation in the SFR active in the disk of M51 from one side to the other apparently as a direct result of the passage of NGC 5195.

\subsection{Next Steps in the Analysis of the System}

While the human eye is an ideal detector for picking up patterns, the human brain is prone to bias.  Thus our method of visual inspection for association selection and turnoff magnitude is prone to the bias of the observer.  In order to reduce this effect, we plan to supplement our visual inspection methods by using an automated selection algorithm, to produce an initial set of candidate associations and ages, and then using visual inspection of each candidate to eliminate false positives and assure accurate age fits.

Also, it has been shown that fitting model SEDs to optical broadband colors alone can be inaccurate, in that it is difficult to differentiate between the reddening effects of age, dust, and metallicity (Cervino \& Luridiana 2006, Anders et al. 2004, Piskunov 2009).  As such, in the next paper we will incoorporate UV data from GALEX and IR data from Spitzer, in order to better differentiate the reddening effects of stellar population age, dust content, and metallicity.

In the future, we plan to apply these same techniques to stellar associations in M83 and NGC 4214, using the multiwavelength images of this galaxy taken with the new Hubble Space Telescope Wide Field Camera 3 (WFC3).  The wide range of wavelength sensitivity available with WFC3 will allow for better differentiation between the reddening effects of age, dust, and metallicity, and thus more accurate dating of stellar associations.  Findings from the present study of M51 will be compared and contrasted with those from the non-interacting spiral galaxy M83.

\section{Conclusions}

We have taken the first high-reslution, multiwavelength, system-wide imaging dataset of the interacting spiral system M51/NGC 5195 and conducted a photometric study of the stellar associations in the system in an effort to measure and characterize the SF in the system, using the associations as age tracers.  We identified 969 stellar groupings of interest, and from this set drew a smaller sample of 120 associations that were deemed potential candidates for further analysis.  Those associations which were isolated enough to be unequivocally regarded as coeval and largely unpolluted by field stars were included in a final sample explored in this paper.  This subset was extracted and presented as the best representives the properties of the stellar association population resident in M51.

To summarize, from the selection, photmetry and SED fitting of 120 single-aged stellar associations in the M51/NGC 5195 system, we conclude the following main points:
\begin{enumerate}
\item{Visual inspection of the CM and CC diagrams for each potential stellar association is an effective way to weed out groups of star that are not coeval.  We find that the vast majority of the single-aged stellar associations selected in this manner lie along one of the spiral arms on M51, or on spurs coming off the spiral arms.}
\item{Very few single-aged associations are found in the interarm regions, suggesting that either the conditions in these inter-arm regions are not ripe for the formation of stellar associations, or that stellar associations fromed here disperse quickly after their birth.}
\item{There is an enhancement in the number of single-aged stellar associations in the northern arm spiral arm closest to the companion, most likely triggered by the intereaction of M51 and NGC 5195.}
\item{No correlation between position within the galaxy and age of the association was found.  There is also not an apparent relationship between the physical size of the association and position.}
\item{We see few instances of sequential formation of stellar associations, which suggests that this type of triggering may only occur on scales smaller than investigated in this study (i.e. smaller $\sim$5 pc, on the scale of compact clusters).}
\item{Single-aged associations in this system range in size from 5 to 90 parsecs, with the peak in the size-number density histogram occurring at an average radius of $\sim$25 parsecs, hinting that this may be the dominant size of the scale length for collapse within the molecular clouds that stellar associations form from.}
\item{There is a slight trend for more massive associations to to exist perferentially in the region of M51 closest to NGC 5195.  This indicates that the galaxy interaction facilitates the collapse of larger segments of molecular hygrogen clouds than would occur without this interaction, resulting in higher-mass stellar associations.}
\item{Even with the resolution of Hubble ACS/WFC, main sequence turnoffs are not evident for the vast majority of stellar associations in M51, due to the limiting magnitude of our images, and the vertical nature of the turnoff at the bright end of the main sequence.}
\end{enumerate}

This type of study has not been possible before because no other face-on spiral has been completely imaged in the photometric bands necessary to sufficient depth and resolution to conduct such an investigation on an entire external galaxy.  In recognition of this fact, the study itself has been very challenging and while we have had some success in measuring the ages of associations in our sample, there have been many fundamental limits to what we can achieve and access because of the distance to the galaxy, the limiting magnitude of the datasets, and the available bandwidths, even when done with the exquisite optics of the HST from space. 

Despite these limitations, we do believe the dataset provides an unprecedented view of the very recent SF process across the entire disk of the galaxy and reflects a definitive change in the SFR from one side of the galaxy to the other probably as a direct result of the interaction with NGC 5195.  This paper represents only the first part of a larger study to study the various SF complexes in M51 using multiwavelength datasets including additional data from GALEX and Spitzer.  That work will be reported in subsequent publications.

\vspace{-\baselineskip}
\acknowledgments{We acknowledge financial support of this work by NASA/STScI Grant HST-AR-10684.01-A.}

\clearpage
\begin{deluxetable*}{llllllllllll}

\tablecaption{Stellar Association Parameters for 120 Single-Aged Associations in the M51/NGC 5195 System}

\tabletypesize{\scriptsize}
\tablewidth{0pt}

\tablehead{\colhead{Identifier} & \colhead{RA\tablenotemark{a}} & \colhead{Dec.\tablenotemark{a}} & \colhead{$M_{B_{\hbox{\tiny int}}}$\tablenotemark{c}} & \colhead{$\!\!\!\!$M$_{V_{\hbox{\tiny int}}}$\tablenotemark{c}} & \colhead{$\!\!\!\!$M$_{I_{\hbox{\tiny int}}}$\tablenotemark{c}} & \colhead{$\!\!\!\!$M$_{B_ {\ast}}$\tablenotemark{d}} & \colhead{M$_{V_{\ast}}$\tablenotemark{d}} & \colhead{M$_{I_{\ast}}$\tablenotemark{d}} & \colhead{$\!\!\!\!$Radius} & \colhead{$\!\!\!\!$Mass} & \colhead{$\!\!\!\!$Age} \\
 & & & & & & & & & \colhead{$\!\!\!\!$(pc)\tablenotemark{e}} & \colhead{$\!\!\!\!$(M$_{\odot}$)} & \colhead{$\!\!\!$(Myrs)} }

\startdata
c1022x3960y1058x4062y1147 & 13:30:6.6247 & 47:14:33.358 & -9.42 & -9.51 & -10.0 & -8.19 & -8.25 & -8.55 & 61.0 & 1.37$\times$10$^{5}$ & 86 \\
c3021x450y731x488y760 & 13:29:58.297 & 47:14:11.042 & -10.2 & -10.3 & -11.0 & -9.64 & -9.82 & -10.4 & 29.4 & 5.05$\times$10$^{5}$ & 200 \\
c3021x489y601x526y630 & 13:29:58.897 & 47:14:9.6002 & -10.0 & -10.1 & -10.5 & -8.91 & -9.05 & -9.23 & 21.4 & 2.41$\times$10$^{5}$ & 86 \\
c3021x708y1939x758y1988 & 13:29:52.393 & 47:13:55.922 & -10.0 & -10.1 & -10.5 & -8.97 & -9.11 & -9.47 & 36.0 & 2.47$\times$10$^{5}$ & 86 \\
c3021x782y1340x816y1376 & 13:29:55.294 & 47:13:53.766 & -8.93 & -9.00 & -9.41 & -7.93 & -8.03 & -8.19 & 26.4 & 8.43$\times$10$^{4}$ & 86 \\
c3021x889y657x923y688 & 13:29:58.535 & 47:13:49.797 & -10.6 & -10.5 & -10.4 & -9.89 & -9.81 & -9.69 & 24.1 & 2.34$\times$10$^{4}$ & 4 \\
c3021x1124y1635x1143y1662 & 13:29:53.785 & 47:13:36.833 & -9.12 & -9.00 & -8.96 & -8.24 & -8.22 & -8.30 & 22.8 & 5.61$\times$10$^{3}$ & 4 \\
c3021x1154y1610x1177y1651 & 13:29:53.902 & 47:13:35.405 & -8.24 & -8.22 & -8.76 & -7.17 & -7.24 & -7.88 & 33.5 & 3.31$\times$10$^{4}$ & 57 \\
c3021x1508y293x1566y332 & 13:30:0.0732 & 47:13:19.557 & -9.72 & -9.80 & -10.3 & -8.69 & -8.94 & -9.56 & 39.7 & 2.40$\times$10$^{5}$ & 140 \\
c3021x1533y136x1596y190 & 13:30:0.7910 & 47:13:18.115 & -10.5 & -10.4 & -10.9 & -9.95 & -9.97 & -10.2 & 55.0 & 2.59$\times$10$^{4}$ & 5 \\
c3021x2008y1044x2042y1071 & 13:29:56.400 & 47:12:54.000 & -8.28 & -8.21 & -8.19 & -7.66 & -7.59 & -7.38 & 25.0 & 2.62$\times$10$^{3}$ & 4 \\
c3021x2016y1329x2042y1352 & 13:29:55.056 & 47:12:53.286 & -8.22 & -8.21 & -8.57 & -7.59 & -7.61 & -7.71 & 13.8 & 3.09$\times$10$^{3}$ & 5 \\
c3021x2031y1397x2057y1430 & 13:29:54.697 & 47:12:52.558 & -9.07 & -9.05 & -9.35 & -8.11 & -8.13 & -8.12 & 18.5 & 6.68$\times$10$^{3}$ & 5 \\
c3021x2122y1118x2145y1141 & 13:29:56.015 & 47:12:48.603 & -8.06 & -8.00 & -8.09 & -7.62 & -7.54 & -7.47 & 14.2 & 2.17$\times$10$^{3}$ & 4 \\
c3021x2847y1871x2878y1907 & 13:29:52.199 & 47:12:11.524 & -9.53 & -9.52 & -9.48 & -8.82 & -8.82 & -8.68 & 23.0 & 8.36$\times$10$^{3}$ & 4 \\
c3021x3532y20x3577y56 & 13:30:0.8642 & 47:11:40.556 & -10.4 & -10.4 & -10.7 & -9.38 & -9.37 & -9.56 & 30.3 & 2.40$\times$10$^{4}$ & 5 \\
c3021x3823y1487x3857y1521 & 13:29:53.833 & 47:11:23.995 & -8.30 & -8.19 & -8.65 & -7.19 & -7.31 & -7.65 & 27.7 & 3.29$\times$10$^{3}$ & 5 \\
c3021x3842y1462x3880y1493 & 13:29:53.928 & 47:11:22.923 & -8.15 & -8.21 & -8.92 & -6.67 & -6.81 & -7.04 & 36.7 & 4.91$\times$10$^{4}$ & 110 \\
c3021x3898y716x3935y757 & 13:29:57.454 & 47:11:21.481 & -8.95 & -9.00 & -9.37 & -8.10 & -8.18 & -8.29 & 31.0 & 6.20$\times$10$^{3}$ & 5 \\
c3022x517y1929x543y1960 & 13:30:2.5195 & 47:14:10.685 & -7.26 & -7.32 & -7.77 & -5.82 & -6.06 & -6.38 & 14.5 & 1.82$\times$10$^{4}$ & 86 \\
c3022x628y1633x683y1680 & 13:30:3.9111 & 47:14:4.9172 & -7.89 & -7.85 & -8.13 & -6.51 & -6.53 & -6.54 & 24.7 & 2.24$\times$10$^{3}$ & 5 \\
c3022x904y1815x949y1860 & 13:30:2.9772 & 47:13:50.525 & -8.04 & -8.15 & -8.74 & -6.49 & -6.72 & -6.93 & 28.1 & 5.59$\times$10$^{4}$ & 160 \\
c3022x1022y18291058y1866 & 13:30:2.8784 & 47:13:45.114 & -8.55 & -8.61 & -9.15 & -7.58 & -7.64 & -7.98 & 22.5 & 6.08$\times$10$^{4}$ & 86 \\
c3022x1186y571x1205y590 & 13:30:9.1186 & 47:13:39.360 & -7.16 & -7.14 & -7.41 & -6.72 & -6.76 & -6.73 & 20.3 & 1.14$\times$10$^{3}$ & 5 \\
c3022x1435y1792x1470y1826 & 13:30:2.9992 & 47:13:24.597 & -7.93 & -8.09 & -8.59 & -6.88 & -7.00 & -7.25 & 22.6 & 5.05$\times$10$^{4}$ & 160 \\
c3022x1521y1530x1555y1564 & 13:30:4.2736 & 47:13:20.999 & -8.86 & -8.75 & -8.86 & -8.05 & -8.05 & -8.06 & 31.9 & 4.49$\times$10$^{3}$ & 4 \\
c3022x1633y721x1664y746 & 13:30:8.2800 & 47:13:16.673 & -7.87 & -8.18 & -8.86 & -7.25 & -7.52 & -7.80 & 32.3 & 8.80$\times$10$^{4}$ & 310 \\
c3022x1680y1247x1705y1277 & 13:30:5.6396 & 47:13:13.446 & -8.83 & -8.85 & -9.43 & -8.04 & -8.15 & -8.71 & 15.3 & 7.82$\times$10$^{4}$ & 86 \\
c3022x1708y820x1723y844 & 13:30:7.7526 & 47:13:13.075 & -8.28 & -8.29 & -9.20 & -7.88 & -7.85 & -8.42 & 16.2 & 2.15$\times$10$^{4}$ & 22 \\
c3022x1842y1441x1868y1463 & 13:30:4.6801 & 47:13:5.1654 & -9.26 & -9.21 & -9.77 & -8.23 & -8.20 & -8.57 & 11.1 & 8.27$\times$10$^{3}$ & 5 \\
c3022x1875y1444x1915y1469 & 13:30:4.6545 & 47:13:2.9956 & -8.78 & -8.87 & -9.43 & -8.01 & -8.14 & -8.31 & 26.7 & 7.66$\times$10$^{4}$ & 86 \\
c3022x1965y1585x2005y1614 & 13:30:3.9367 & 47:12:58.683 & -8.08 & -8.18 & -9.05 & -6.92 & -7.10 & -7.79 & 17.0 & 8.30$\times$10$^{4}$ & 250 \\
c3022x2085y1165x2106y1188 & 13:30:5.9985 & 47:12:53.643 & -7.90 & -7.93 & -8.37 & -7.42 & -7.42 & -7.65 & 28.0 & 2.37$\times$10$^{3}$ & 5 \\
c3022x2101y588x2126y614 & 13:30:8.8073 & 47:12:53.286 & -7.50 & -7.57 & -7.96 & -6.75 & -6.80 & -6.95 & 16.9 & 1.65$\times$10$^{3}$ & 5 \\
c3022x2309y1481x2332y1505 & 13:30:4.3688 & 47:12:42.121 & -7.13 & -7.12 & -7.30 & -6.45 & -6.53 & -6.72 & 27.4 & 1.11$\times$10$^{3}$ & 5 \\
c3022x2324y1687x2354y1731 & 13:30:3.3105 & 47:12:40.679 & -8.31 & -8.37 & -8.94 & -7.38 & -7.55 & -8.00 & 27.2 & 4.91$\times$10$^{4}$ & 86 \\
c3022x2449y1572x2491y1611 & 13:30:3.8891 & 47:12:34.554 & -9.71 & -9.85 & -10.2 & -9.01 & -9.22 & -9.47 & 43.7 & 1.76$\times$10$^{5}$ & 86 \\
c3022x2480y1085x2514y1112 & 13:30:6.2878 & 47:12:33.483 & -7.65 & -7.90 & -8.83 & -6.54 & -6.84 & -7.65 & 32.1 & 9.92$\times$10$^{4}$ & 430 \\
c3022x2618y1453x2655y1485 & 13:30:4.4384 & 47:12:26.273 & -10.0 & -9.99 & -10.0 & -9.41 & -9.37 & -9.36 & 37.1 & 1.39$\times$10$^{4}$ & 4 \\
c3022x2629y959x2659y986 & 13:30:6.8884 & 47:12:26.644 & -9.72 & -9.63 & -9.63 & -8.99 & -8.89 & -8.71 & 21.8 & 9.92$\times$10$^{3}$ & 4 \\
c3022x2709y1546x2723y1568 & 13:30:3.9843 & 47:12:22.318 & -7.50 & -7.41 & -7.75 & -7.12 & -7.03 & -7.31 & 5.40 & 1.54$\times$10$^{3}$ & 5 \\
c3022x2735y1709x2753y1728 & 13:30:3.1677 & 47:12:20.519 & -7.27 & -7.33 & -7.76 & -6.60 & -6.68 & -7.07 & 17.6 & 1.84$\times$10$^{4}$ & 86 \\
c3022x3101y1813x3143y1850 & 13:30:2.5671 & 47:12:1.4447 & -8.82 & -8.92 & -9.35 & -8.08 & -8.22 & -8.52 & 40.0 & 7.75$\times$10$^{4}$ & 86 \\
c3022x3561y786x3605y846 & 13:30:7.4890 & 47:11:39.842 & -8.96 & -9.28 & -10.3 & -8.08 & -8.51 & -9.55 & 47.5 & 4.34$\times$10$^{5}$ & 540 \\
c3022x3861y1479x3894y1527 & 13:30:3.9843 & 47:11:24.365 & -7.62 & -7.78 & -8.72 & -6.58 & -6.67 & -7.51 & 25.2 & 8.43$\times$10$^{4}$ & 390 \\
c3022x3861y936x3908y973 & 13:30:6.7456 & 47:11:24.722 & -9.51 & -9.47 & -9.43 & -8.79 & -8.77 & -8.68 & 33.0 & 8.26$\times$10$^{3}$ & 4 \\
c4021x1738y32x1775y74 & 13:29:41.953 & 47:13:1.1965 & -8.01 & -8.04 & -8.70 & -7.20 & -7.30 & -7.38 & 33.0 & 3.96$\times$10$^{4}$ & 96 \\
c4021x2677y615x2717y650 & 13:29:38.902 & 47:12:13.680 & -8.69 & -8.72 & -8.91 & -8.14 & -8.17 & -8.19 & 25.5 & 4.70$\times$10$^{3}$ & 5 \\
c4021x2711y385x2759y443 & 13:29:39.960 & 47:12:12.238 & -9.14 & -9.27 & -9.67 & -8.60 & -8.73 & -8.99 & 41.8 & 1.04$\times$10$^{5}$ & 86 \\
c4021x3433y239x3487y286 & 13:29:40.462 & 47:11:36.958 & -8.08 & -8.28 & -9.43 & -7.36 & -7.63 & -8.64 & 54.4 & 6.20$\times$10$^{3}$ & 7 \\
c4021x3464y32x3497y67 & 13:29:41.495 & 47:11:36.244 & -9.13 & -9.19 & -9.81 & -8.34 & -8.44 & -8.78 & 20.0 & 9.77$\times$10$^{4}$ & 77 \\
c4021x3501y776x3544y818 & 13:29:37.921 & 47:11:33.003 & -8.80 & -8.90 & -9.59 & -7.45 & -7.64 & -8.15 & 28.6 & 9.79$\times$10$^{4}$ & 130 \\
c4021x3707y534x3744y571 & 13:29:39.023 & 47:11:23.280 & -8.00 & -7.85 & -7.91 & -7.16 & -7.06 & -6.90 & 22.8 & 2.00$\times$10$^{3}$ & 4 \\
c4022x186y169x217y197 & 13:29:51.936 & 47:14:21.479 & -7.90 & -7.94 & -8.30 & -7.17 & -7.19 & -7.16 & 25.5 & 2.34$\times$10$^{3}$ & 5 \\
c4022x587y467x642y497 & 13:29:50.375 & 47:14:0.6051 & -9.06 & -9.13 & -9.28 & -8.26 & -8.31 & -8.23 & 44.4 & 6.65$\times$10$^{3}$ & 5 \\
c4022x692y1002x744y1066 & 13:29:47.640 & 47:13:54.480 & -8.65 & -8.75 & -9.25 & -7.54 & -7.71 & -7.93 & 55.6 & 6.72$\times$10$^{4}$ & 86 \\
c4022x873y435x910y474 & 13:29:50.471 & 47:13:46.556 & -8.00 & -8.50 & -9.42 & -6.90 & -7.45 & -8.26 & 36.3 & 2.07$\times$10$^{5}$ & 610 \\
c4022x880y668x911y698 & 13:29:49.321 & 47:13:46.199 & -8.20 & -8.27 & -8.61 & -7.59 & -7.77 & -7.98 & 19.9 & 3.10$\times$10$^{3}$ & 5 \\
c4022x1032y1210x1081y1238 & 13:29:46.633 & 47:13:37.561 & -9.44 & -9.39 & -9.48 & -8.55 & -8.52 & -8.57 & 33.9 & 7.79$\times$10$^{3}$ & 4 \\
c4022x1093y1186x1115y1208 & 13:29:46.750 & 47:13:35.034 & -8.94 & -9.07 & -9.58 & -8.21 & -8.48 & -8.80 & 22.4 & 1.27$\times$10$^{5}$ & 160 \\
c4022x1158y346x1180y376 & 13:29:50.855 & 47:13:32.878 & -7.71 & -7.88 & -8.79 & -6.73 & -6.88 & -7.44 & 25.7 & 9.14$\times$10$^{4}$ & 390 \\
c4022x1713y118x1738y161 & 13:29:51.793 & 47:13:5.5224 & -8.72 & -8.66 & -8.76 & -8.41 & -8.37 & -8.32 & 22.6 & 3.98$\times$10$^{3}$ & 4 \\
c4022x1775y105x1807y139 & 13:29:51.936 & 47:13:2.2814 & -8.08 & -8.15 & -9.04 & -7.32 & -7.44 & -7.68 & 35.4 & 8.22$\times$10$^{4}$ & 250 \\
\enddata
\end{deluxetable*}

\clearpage
\addtocounter{table}{-1}
\begin{deluxetable*}{llllllllllll}

\tablecaption{(continued)}

\tabletypesize{\scriptsize}
\tablewidth{0pt}

\tablehead{\colhead{Identifier} & \colhead{RA\tablenotemark{a}} & \colhead{Dec.\tablenotemark{a}} & \colhead{$M_{B_{\hbox{\tiny int}}}$\tablenotemark{c}} & \colhead{$\!\!\!\!$M$_{V_{\hbox{\tiny int}}}$\tablenotemark{c}} & \colhead{$\!\!\!\!$M$_{I_{\hbox{\tiny int}}}$\tablenotemark{c}} & \colhead{$\!\!\!\!$M$_{B_ {\ast}}$\tablenotemark{d}} & \colhead{M$_{V_{\ast}}$\tablenotemark{d}} & \colhead{M$_{I_{\ast}}$\tablenotemark{d}} & \colhead{$\!\!\!\!$Radius} & \colhead{$\!\!\!\!$Mass} & \colhead{$\!\!\!\!$Age} \\
 & & & & & & & & & \colhead{$\!\!\!\!$(pc)\tablenotemark{e}} & \colhead{$\!\!\!\!$(M$_{\odot}$)} & \colhead{$\!\!\!$(Myrs)} }
%\tablecolumns{12}
%\section{}
%

\startdata
c4022x1965y456x1995y487 & 13:29:50.185 & 47:12:52.558 & -8.48 & -8.44 & -8.39 & -8.00 & -7.96 & -7.78 & 25.5 & 3.17$\times$10$^{3}$ & 4 \\
c4022x2109y746x2156y785 & 13:29:48.647 & 47:12:44.277 & -8.09 & -8.23 & -9.03 & -7.41 & -7.63 & -8.02 & 41.5 & 7.67$\times$10$^{4}$ & 220 \\
c4022x2149y1043x2199y1084 & 13:29:47.182 & 47:12:41.764 & -8.56 & -8.76 & -9.54 & -7.94 & -8.16 & -8.62 & 53.4 & 1.32$\times$10$^{5}$ & 250 \\
c4022x2289y1064x2329y1103 & 13:29:47.065 & 47:12:35.282 & -8.01 & -8.15 & -8.87 & -7.01 & -7.21 & -7.63 & 31.3 & 6.48$\times$10$^{4}$ & 200 \\
c4022x2457y1512x2491y1553 & 13:29:44.831 & 47:12:26.273 & -9.61 & -9.58 & -9.74 & -8.81 & -8.77 & -8.71 & 35.2 & 9.17$\times$10$^{3}$ & 4 \\
c4022x2554y502x2612y566 & 13:29:49.727 & 47:12:22.318 & -9.34 & -9.39 & -10.0 & -8.59 & -8.63 & -9.00 & 36.4 & 1.46$\times$10$^{5}$ & 110 \\
c4022x2787y25x2821y58 & 13:29:52.104 & 47:12:12.238 & -9.43 & -9.40 & -9.35 & -8.78 & -8.76 & -8.67 & 24.5 & 7.59$\times$10$^{3}$ & 4 \\
c4022x2826y1550x2879y1603 & 13:29:44.520 & 47:12:7.1987 & -7.99 & -8.03 & -8.84 & -6.64 & -6.70 & -7.03 & 50.7 & 1.95$\times$10$^{4}$ & 31 \\
c4022x2840y571x2883y602 & 13:29:49.390 & 47:12:8.2836 & -8.94 & -9.03 & -9.63 & -8.33 & -8.55 & -9.15 & 32.7 & 1.18$\times$10$^{5}$ & 140 \\
c4022x2945y717x2974y749 & 13:29:48.673 & 47:12:3.2437 & -8.12 & -8.04 & -8.74 & -7.48 & -7.49 & -7.79 & 13.8 & 2.39$\times$10$^{4}$ & 40 \\
c4022x3333y1755x3350y1779 & 13:29:43.487 & 47:11:43.083 & -8.26 & -8.29 & -8.57 & -7.40 & -7.50 & -7.76 & 11.7 & 3.21$\times$10$^{3}$ & 5 \\
c4022x3437y1043x3468y1084 & 13:29:46.944 & 47:11:38.400 & -7.96 & -7.82 & -8.59 & -7.48 & -7.37 & -7.53 & 38.6 & 1.57$\times$10$^{4}$ & 24 \\
c4022x3868y340x3912y396 & 13:29:50.302 & 47:11:17.526 & -9.25 & -9.36 & -10.0 & -8.44 & -8.67 & -9.30 & 46.2 & 1.85$\times$10$^{5}$ & 180 \\
c5021x264y445x314y509 & 13:29:58.967 & 47:11:5.6341 & -11.3 & -11.2 & -11.4 & -10.5 & -10.5 & -10.6 & 41.6 & 4.56$\times$10$^{4}$ & 4 \\
c5021x665y1627x700y1662 & 13:29:53.280 & 47:10:44.046 & -10.8 & -10.9 & -11.1 & -10.0 & -10.1 & -10.3 & 24.3 & 3.49$\times$10$^{4}$ & 5 \\
c5021x669y1626x701y1667 & 13:29:53.280 & 47:10:44.046 & -10.8 & -10.9 & -11.1 & -10.2 & -10.2 & -10.4 & 32.0 & 3.61$\times$10$^{4}$ & 5 \\
c5021x703y849x726y885 & 13:29:56.975 & 47:10:43.675 & -7.73 & -8.00 & -9.13 & -6.63 & -7.11 & -8.02 & 21.0 & 1.40$\times$10$^{5}$ & 540 \\
c5021x739y1142x770y1180 & 13:29:55.583 & 47:10:41.519 & -8.33 & -8.32 & -8.86 & -7.76 & -7.80 & -7.94 & 29.1 & 3.61$\times$10$^{4}$ & 57 \\
c5021x791y1253x813y1271 & 13:29:55.081 & 47:10:38.635 & -7.31 & -7.44 & -8.16 & -6.50 & -6.67 & -6.98 & 9.30 & 3.38$\times$10$^{4}$ & 200 \\
c5021x791y1620x828y1644 & 13:29:53.327 & 47:10:37.921 & -8.88 & -8.96 & -9.49 & -7.93 & -8.19 & -8.66 & 34.5 & 8.27$\times$10$^{4}$ & 86 \\
c5021x1141y1883x1167y1934 & 13:29:51.936 & 47:10:20.274 & -8.49 & -8.49 & -9.10 & -7.40 & -7.43 & -7.80 & 37.2 & 4.25$\times$10$^{4}$ & 57 \\
c5021x1149y1574x1182y1604 & 13:29:53.422 & 47:10:20.274 & -9.58 & -9.58 & -10.1 & -8.72 & -8.75 & -9.06 & 29.0 & 1.14$\times$10$^{5}$ & 57 \\
c5021x1169y1324x1217y1377 & 13:29:54.576 & 47:10:19.560 & -10.5 & -10.5 & -11.1 & -9.68 & -9.77 & -10.2 & 35.8 & 2.83$\times$10$^{5}$ & 57 \\
c5021x1178y1641x1213y1673 & 13:29:53.111 & 47:10:18.846 & -9.10 & -9.07 & -9.77 & -8.32 & -8.34 & -8.78 & 37.8 & 5.79$\times$10$^{4}$ & 38 \\
c5021x1253y1509x1284y1540 & 13:29:53.712 & 47:10:15.605 & -9.06 & -8.90 & -8.73 & -8.44 & -8.33 & -8.03 & 28.8 & 5.20$\times$10$^{3}$ & 4 \\
c5021x1735y1344x1775y1378 & 13:29:54.360 & 47:9:51.833 & -8.35 & -8.41 & -8.87 & -7.42 & -7.47 & -7.66 & 35.6 & 4.97$\times$10$^{4}$ & 86 \\
c5021x1852y874x1899y907 & 13:29:56.568 & 47:9:46.793 & -9.22 & -9.33 & -9.86 & -8.60 & -8.67 & -8.98 & 32.3 & 1.63$\times$10$^{5}$ & 160 \\
c5021x2324y1077x2369y1134 & 13:29:55.440 & 47:9:23.035 & -8.62 & -8.69 & -9.32 & -7.75 & -7.78 & -8.01 & 50.0 & 7.00$\times$10$^{4}$ & 96 \\
c5021x2429y1611x2449y1636 & 13:29:52.943 & 47:9:17.638 & -8.15 & -8.02 & -7.95 & -7.66 & -7.57 & -7.36 & 14.3 & 2.28$\times$10$^{3}$ & 4 \\
c5021x2430y121x2470y157 & 13:30:0.0256 & 47:9:19.794 & -8.83 & -8.83 & -9.07 & -8.04 & -8.10 & -8.19 & 35.3 & 5.33$\times$10$^{3}$ & 5 \\
c5021x2537y414x2560y438 & 13:29:58.630 & 47:9:14.397 & -8.61 & -8.60 & -8.79 & -8.02 & -8.05 & -8.23 & 22.6 & 4.27$\times$10$^{3}$ & 5 \\
c5021x2552y318x2590y354 & 13:29:59.040 & 47:9:13.326 & -8.90 & -8.90 & -9.14 & -8.03 & -8.04 & -8.11 & 34.4 & 5.66$\times$10$^{3}$ & 5 \\
c5021x2806y1020x2840y1051 & 13:29:55.631 & 47:9:0.0054 & -8.51 & -8.36 & -8.31 & -7.80 & -7.67 & -7.41 & 26.5 & 3.18$\times$10$^{3}$ & 4 \\
c5021x2854y667x2890y702 & 13:29:57.286 & 47:8:57.835 & -9.73 & -9.66 & -9.61 & -9.28 & -9.24 & -9.13 & 25.0 & 1.00$\times$10$^{4}$ & 4 \\
c5021x3033y439x3063y481 & 13:29:58.344 & 47:8:49.925 & -8.38 & -8.39 & -8.64 & -7.72 & -7.78 & -7.99 & 43.4 & 3.53$\times$10$^{3}$ & 5 \\
c5021x3054y1592x3074y1629 & 13:29:52.822 & 47:8:47.041 & -8.54 & -8.57 & -9.09 & -7.74 & -7.86 & -8.68 & 28.9 & 5.91$\times$10$^{4}$ & 86 \\
c5021x3454y1722x3527y1797 & 13:29:52.031 & 47:8:25.796 & -9.07 & -9.19 & -9.77 & -8.24 & -8.29 & -8.64 & 55.8 & 1.45$\times$10$^{5}$ & 160 \\
c5022x754y1102x784y1132 & 13:30:5.8813 & 47:10:44.760 & -8.16 & -8.50 & -9.43 & -7.39 & -7.78 & -8.45 & 26.0 & 1.82$\times$10$^{5}$ & 490 \\
c5022x770y887x805y922 & 13:30:6.9104 & 47:10:44.403 & -9.10 & -9.24 & -9.72 & -8.33 & -8.51 & -8.99 & 23.8 & 1.46$\times$10$^{5}$ & 160 \\
c5022x997y1153x1075y1223 & 13:30:5.4711 & 47:10:31.439 & -9.61 & -9.79 & -10.6 & -8.72 & -8.97 & -9.47 & 74.7 & 3.86$\times$10$^{5}$ & 280 \\
c5022x1158y1106x1236y1208 & 13:30:5.5664 & 47:10:23.158 & -10.3 & -10.4 & -10.9 & -9.60 & -9.78 & -10.1 & 90.7 & 4.53$\times$10$^{5}$ & 160 \\
c5022x1240y1176x1269y1207 & 13:30:5.4235 & 47:10:20.274 & -9.55 & -9.50 & -9.66 & -8.95 & -8.94 & -9.06 & 28.9 & 8.73$\times$10$^{3}$ & 4 \\
c5022x1771y1568x1806y1607 & 13:30:3.3361 & 47:9:53.275 & -10.3 & -10.2 & -10.2 & -9.86 & -9.77 & -9.73 & 25.5 & 1.75$\times$10$^{4}$ & 4 \\
c5022x1994y1499x2037y1542 & 13:30:3.6950 & 47:9:42.124 & -8.90 & -8.89 & -8.96 & -7.78 & -7.88 & -8.04 & 45.6 & 4.76$\times$10$^{3}$ & 4 \\
c5022x1997y1528x2043y1581 & 13:30:3.4790 & 47:9:41.753 & -9.80 & -9.91 & -10.4 & -8.99 & -9.15 & -9.53 & 47.7 & 1.95$\times$10$^{5}$ & 86 \\
c5022x2754y1933x2787y1967 & 13:30:1.3916 & 47:9:3.6035 & -9.20 & -9.19 & -9.56 & -8.53 & -8.56 & -8.84 & 24.8 & 7.64$\times$10$^{3}$ & 5 \\
c5022x3634y1749x3685y1816 & 13:30:2.0397 & 47:8:20.042 & -9.27 & -9.19 & -9.16 & -8.55 & -8.48 & -8.38 & 44.6 & 6.52$\times$10$^{3}$ & 4 \\
c6021x1727y965x1790y1024 & 13:29:36.793 & 47:9:44.280 & -9.60 & -9.61 & -10.2 & -8.95 & -9.02 & -9.58 & 66.5 & 1.19$\times$10$^{5}$ & 57 \\
c6021x1921y1183x1954y1216 & 13:29:35.760 & 47:9:35.285 & -7.54 & -7.66 & -8.40 & -6.76 & -6.85 & -7.19 & 37.7 & 4.21$\times$10$^{4}$ & 200 \\
c6021x3007y329x3096y394 & 13:29:39.481 & 47:8:42.001 & -9.95 & -10.1 & -11.0 & -9.01 & -9.24 & -9.81 & 66.1 & 7.40$\times$10$^{5}$ & 390 \\
c6022$\times$10y1270x53y1333 & 13:29:45.816 & 47:11:13.200 & -9.31 & -9.34 & -9.91 & -8.59 & -8.72 & -9.07 & 57.3 & 1.22$\times$10$^{5}$ & 86 \\
c6022x402y1882x442y1909 & 13:29:42.791 & 47:10:52.684 & -10.1 & -10.0 & -9.95 & -9.43 & -9.45 & -9.30 & 30.8 & 1.40$\times$10$^{4}$ & 4 \\
c6022x433y325x461y367 & 13:29:50.423 & 47:10:53.755 & -9.49 & -9.52 & -9.90 & -8.70 & -8.77 & -9.07 & 34.3 & 1.01$\times$10$^{4}$ & 5 \\
c6022x669y359x695y390 & 13:29:50.255 & 47:10:42.233 & -10.9 & -10.9 & -10.9 & -10.1 & -10.1 & -10.0 & 26.1 & 3.18$\times$10$^{4}$ & 4 \\
c6022x957y1773x1014y1824 & 13:29:43.150 & 47:10:24.957 & -10.2 & -10.2 & -10.5 & -9.27 & -9.40 & -9.66 & 46.6 & 1.93$\times$10$^{4}$ & 5 \\
c6022x1313y77x1352y130 & 13:29:51.456 & 47:10:9.8373 & -9.40 & -9.36 & -9.85 & -8.55 & -8.61 & -8.94 & 32.4 & 9.30$\times$10$^{3}$ & 5 \\
c6022x1450y1665x1480y1688 & 13:29:43.656 & 47:10:1.1993 & -7.90 & -8.15 & -9.20 & -7.27 & -7.55 & -8.26 & 22.2 & 1.42$\times$10$^{5}$ & 490 \\
c6022x1887y953x1912y972 & 13:29:47.113 & 47:9:40.682 & -9.59 & -9.51 & -9.64 & -9.13 & -9.06 & -9.14 & 25.2 & 8.96$\times$10$^{3}$ & 4 \\
\enddata

\tablenotetext{a}{Epoch J2000}
\tablenotetext{b}{All magnitudes in the Vega magnitude system}
\tablenotetext{c}{Total integrated magnitude from a circular aperture of radius and central position listed in Table 1.}
\tablenotetext{d}{Total magnitude from summing all sources in the association.  This value will not include diffuse light from unresolved stars, and is thus a lower limit of the magnitude of the association.}
\tablenotetext{e}{The average radius of the association as determined by the average FWHM of the flux density distribution.}
\end{deluxetable*}

\end{document}